\documentclass[12pt,a4paper]{article}
\usepackage{amsmath,amssymb,amsfonts,amsthm,amsopn,graphicx}
\usepackage{tikz} 
\usetikzlibrary{arrows}
\allowdisplaybreaks[3]
\numberwithin{equation}{section} 
\newtheorem{theorem}{Theorem}[section]

\newtheorem{statement}[theorem]{Statement}

\newtheorem{convention}[theorem]{Convention}
\newcommand{\ds}{\displaystyle}
\def\EXP{\textrm{{\large e}}}
\newcommand{\xop}{\boldsymbol{x}}
\newcommand{\pop}{\boldsymbol{p}}
\newcommand{\uop}{\boldsymbol{u}}
\newcommand{\vop}{\boldsymbol{v}}
\newcommand{\ch}{\boldsymbol{\chi}}

\newcommand{\ii}{\mathsf{i}}
\newcommand{\bb}{\mathsf{b}}

\def\EXP{\textrm{{\large e}}}

\renewcommand{\author}[1]{\large\rm #1\\ \bigskip}
\newcommand{\address}[1]{{\normalsize\it #1\\}\bigskip}
\renewcommand{\title}[1]{\bigskip\bigskip\Large\bf #1\bigskip\bigskip\\}

\usepackage[body={15.5cm,22cm}]{geometry}
\begin{document}

\begin{center}

\title{Spectral equations and ground state for the modular $\sinh$-Gordon model.}

\author{Sergey M.~Sergeev.}

\vspace{.4cm}

\address{Department of Theoretical Physics,
         Research School of Physics and Engineering,\\
    Australian National University, Canberra, ACT 0200, Australia\\
    and\\
   Faculty of Science and Technology, \\
   University of Canberra, Bruce ACT 2617, Australia }


\end{center}
\begin{abstract}
We study the modular pair of $TQ$ equations for the quantum $\sin$-Gordon model in the framework of non-compact $\mathcal{U}_{q,q^*}(\widehat{sl}_2)$. We assume some conjectures for the thermodynamic limit allowing one to obtain its ground state.
\end{abstract}



\section{Introduction.}

I this paper we discuss some analytical properties of the Baxter's $Q$-operator for the quantum non-compact sinh-Gordon model 
considered in the framework of the modular double $\mathcal{U}_{q,\overline{q}}(\widehat{sl}_2)$ \cite{Faddeev_1995,Faddeev_1999}.
We do not discuss construction of the model here. 
Instead, we start directly from the quantum curve for it and the pair of $TQ$ equations. 
Quantisation condition \cite{Sergeev_2005,Kashaev_2018,Babelon_2018,Kashaev_2020} is the absence of the poles in the \emph{eigenfunction}  $Q(x)$ except some ``kinematic'' poles coming explicitly from the kernel of the \emph{operator} $Q(x)$ and described by usual quantum dilogarithms. Corresponding spectral equations are given in Section \ref{Section5}, eq. (\ref{BAE}). 
This paper can be seen as a continuation of \cite{Sergeev_2005}. The main purpose of this paper is the study of the spectral equations in the thermodynamic limit. Under certain conjectures we obtain a functional relation for the density of a properly defined Bethe-type variables for the ground state of the model. Our results are in accordance with \cite{Bazhanov_2007,Lukyanov_1997}.

\section{Basic notations.}

Let the self-conjugated pair $\xop,\pop$ form the Heisenberg algebra,
\begin{equation}\label{Heis}
[\xop,\pop]\;=\;\frac{\ii}{2\pi}\;.
\end{equation}
Let also
\begin{equation}\label{bb}
\bb\;=\;\EXP^{\ii\theta}\;,\quad 0 < \theta < \frac{\pi}{2}\;,
\end{equation}
and
\begin{equation}\label{qq}
q\;=\;\EXP^{\ii\pi\bb^2}\;,\quad q^*\;=\;\EXP^{-\ii\pi\bb^{-2}}\;.
\end{equation}

\begin{convention}\label{star-convention}
In what follows, ``star $*$'' will stand for the involution 
\begin{equation}\label{star-involution}
u\;=\;\EXP^{2\pi\bb x}\;\to\; u^*\;=\;\EXP^{2\pi\bb^{-1}x}\;;\quad
q^2\;=\;\EXP^{2\pi\ii\bb^2}\;\to\; q^{\star 2}\;=\;\EXP^{-2\pi\ii\bb^{-2}}\;.
\end{equation}
Involutive antihomomorphism  $*$ is the complex/Hermitian conjugation in the regime of real $x$ in (\ref{star-involution}). However, in general the $*$ involution implies complex $x$ understood as the analytical continuation.
\end{convention}
\noindent

\bigskip

\noindent
Two conjugated Weyl pairs to be introduced are
\begin{equation}\label{uv}
\uop\;=\;\EXP^{2\pi\bb\xop}\;,\quad \vop\;=\;\EXP^{2\pi\bb\pop}\;,\quad \uop\,\vop\;=\;q^2\,\vop\,\uop\;,
\end{equation}
and 
\begin{equation}
\uop^*\;=\;\EXP^{2\pi\bb^{-1}\xop}\;,\quad \vop^*\;=\;\EXP^{2\pi\bb^{-1}\pop}\;,\quad \vop^*\,\uop^*\;=\;q^{*2}\,\uop^*\,\vop^*\;,
\end{equation}
so that
\begin{equation}
[\uop,\vop^*]\;=\;[\vop,\uop^*]\;=\;0\;.
\end{equation}
Dirac's bracket notations are defined by
\begin{equation}\label{bra}
\langle x|\,\uop\;=\;u\,\langle x|\;,\quad 
\langle x|\,\uop^*\;=\;u^*\,\langle x|\;,\quad 
\langle x|\,\vop\;=\;\langle x-\ii\bb|\;,\quad \langle x|\,\vop^*\;=\;\langle x-\ii\bb^{-1}|\;.
\end{equation}

\bigskip

\noindent
In what follows, we will always use
\begin{equation}\label{ux-convention}
u\;\stackrel{def}{=}\;\EXP^{2\pi\bb x}\;,
\quad
u^*\stackrel{def}{=}\;\EXP^{2\pi\bb^{-1}x}\;,
\quad \eta\;=\;\frac{\bb+\bb^{-1}}{2}\;,\quad \sigma\;=\;\frac{\bb-\bb^{-1}}{2\ii}\;.
\end{equation}

\section{Special Functions.}

Now we will define a set of special functions.

\bigskip
\noindent
The $q$-exponent is defined by 
\begin{equation}\label{poh}
(u;q^2)_\infty \;=\; \prod_{n=0}^\infty (1-uq^{2n})\;,
\end{equation}
and shortened $\vartheta$-function -- by
\begin{equation}\label{theta}
\vartheta_1(u)\;=\;(u;q^2)_\infty (q^2u^{-1};q^2)_\infty\;,\quad \vartheta_1(u)\;=\;-\,u\,\vartheta_1(q^2u)\;.
\end{equation}
Jacobi identity reads
\begin{equation}\label{Jacobi}
\frac{\vartheta_1(u)^{}}{\vartheta_1(u)^*}\;=\;
\EXP^{\ii\pi (x+\sigma)^2+\ii\pi c_\bb}\;,\quad c_\bb\;=\;\frac{1}{12}(\bb^2+\bb^{-2})\;.
\end{equation}

\bigskip
\noindent
Two basic $q$-dilogarithmic functions \cite{Faddeev_2001} are:
\begin{equation}\label{phi}
\log\varphi(x)\;=\;\mathop{\int}_{\mathbb{R}+\ii 0} \frac{\ds \EXP^{-2\ii xy}}{\ds 4\sinh(\bb y)\sinh(\bb^{-1}y)}\, \frac{dy}{y}
\;=\; \log\frac{ (-qu;q^2)_\infty^{}}{(-qu;q^2)_\infty^*}\;,
\end{equation}
and \cite{Lukyanov_1997,Bazhanov_2008}
\begin{equation}\label{phi2}
\log\varphi_2(x)\;=\;\mathop{\int}_{\mathbb{R}+\ii 0} \frac{\ds \EXP^{-2\ii xy}}{\ds 8\cosh(2\eta y)\sinh(\bb y)\sinh(\bb^{-1}y)}\, \frac{dy}{y}\;.
\end{equation}

\bigskip

\noindent
Most remarkable properties of $\varphi(x)$ and $\varphi_2(x)$ to be mentioned:
\begin{equation}
\begin{array}{l}
\ds \log\varphi(x-\ii\eta)\,-\,\log\varphi(x+\ii\eta)\;=\;\log(1-u)(1-u^*)\;,\\
[5mm]
\ds \log\varphi_2(x+\ii\eta)\,+\,\log\varphi_2(x-\ii\eta)\;=\;\varphi(x)\;,\\
[5mm]
\ds \log\varphi(x)\,+\,\log\varphi(-x)\;=\;\ii\pi x^2 \,+\, \ii\pi c_\bb\;,\\
[5mm]
\ds \log\varphi_2(x)\,+\,\log\varphi_2(-x)\;=\;\frac{\ii\pi}{2} x^2 \,+\, 2\pi\ii c_\bb\,+\,\frac{\ii\pi}{4}\;.
\end{array}
\end{equation}

\section{$TQ$ -- Equations.}

In this section we will formulate the modular pair of Baxter's $TQ$ equations. In this paper we will not give a detailed formulation of $\mathcal{U}_{q,q^*}(\widehat{sl}_2)$ $\sin$-Gordon model in the terms of the whole machinery of the Quantum Inverse Scattering Method. Instead, our staring point will be corresponding Quantum Characteristic Polynomial.

We start with a few auxiliary notations. 
Let
\begin{equation}\label{AB}
A(u)\;=\;\prod_{\nu=1}^N \left( 1\,+\,q^{-1}\frac{u}{a_\nu}\right)\;,\quad
B(u)\;=\;\prod_{\nu=1}^N \left(1\,+\,q\frac{u}{b_\nu}\right)\;,
\end{equation}
and
\begin{equation}\label{ABprime}
A'(1/u)\;=\;\prod_{\nu=1}^N \left(1\,+\,q\frac{a_\nu}{u}\right)\;,\quad
B'(1/u)\;=\;\prod_{\nu=1}^N \left(1\,+\,q^{-1}\frac{b_\nu}{u}\right)\;.
\end{equation}
Let also
\begin{equation}\label{geomean}
\prod_{\nu=1}^N a_\nu \;=\; a^N\;,\quad
\prod_{\nu=1}^N b_\nu\;=\;b^N\;,
\end{equation}
so that
\begin{equation}
A(u)\;=\;\left(\frac{u}{qa}\right)^N A'(1/u)\;,\quad
B(u)\;=\;\left(\frac{qu}{b}\right)^N B'(1/u)\;.
\end{equation}
We will imply
\begin{equation}\label{params}
a_\nu^{}\;=\;\EXP^{2\pi\bb\alpha_\nu}\;,\quad
a_\nu^{*}\;=\;\EXP^{2\pi\bb^{-1}\alpha_\nu}\;,\quad
b_\nu^{}\;=\;\EXP^{2\pi\bb\beta_\nu}\;,\quad
b_\nu^{*}\;=\;\EXP^{2\pi\bb^{-1}\beta_\nu}\;.
\end{equation}
One more convenient notation:
\begin{equation}
A(u;q^2)_\infty \;=\; \prod_{m=0}^\infty A(q^{2m}u)\;=\;\prod_{\nu=1}^N (-q^{-1}\frac{u}{a_\nu};q^2)_\infty\;,\quad \textrm{etc.}
\end{equation}
Also we will assume for the geometric mean (\ref{geomean})
\begin{equation}\label{abnormalization}
a\;=\;\EXP^{2\pi\bb\mu}\;,\quad b\;=\;\EXP^{-2\pi\bb\mu}\;,\quad 
ab\;=\;1\;.
\end{equation}

\bigskip

\noindent
The Characteristic Polynomial for the $\sin$-Gordon model is
\begin{equation}\label{qcurve}
J(\uop,\vop)\;=\; t^{-1}\,A(\uop) \,\vop\;-\;T(\uop)\;+\;t\,B(\uop)\,\vop^{-1}\;,
\end{equation}
where $A,B$ are defined by (\ref{AB}), $T(\uop)$ is the transfer-matrix,
\begin{equation}\label{transfer}
T(\uop)\;=\;\sum_{n=0}^N T_n \uop^n\;,
\end{equation}
and $N$ is the chain length. Note that here $\uop,\vop$ are the spectral parameters, while
\begin{equation}
T_1,\quad T_2,\quad \dots \quad T_{N-1}
\end{equation}
is the commutative set of the integrals of motion. The opposite terms of the transfer matrix are the ``external'' parameters\footnote{$t$ is an analogue of $q^{S_z}$},
\begin{equation}\label{T0N}
T_0\;=\;t^{-1}+t\;,\quad T_N\;=\;(-)^N (t^{-1}+t)\;.
\end{equation}
Extra condition to be imposed to parameter $t$ (for finite $N$) is
\begin{equation}\label{ts-cond}
|t^2|\,<\,1\;.
\end{equation}
The particular form of $T_0$, $T_N$, $A(u)$ and $B(u)$ follows from the explicit construction of the quantum curve via $L$-operators etc.

\noindent

Characteristic polynomial $J(\uop,\vop)$ must be the normal operator.
Baxter's $TQ$ equations read\footnote{State $|Q\rangle$ is called sometimes $Q_L$.}
\begin{equation}\label{JQ}
J(\uop,\vop)\,|Q\rangle\;=\;J(\uop,\vop)^\dagger |Q\rangle\;=\;0\;.
\end{equation}
In the coordinate representation
\begin{equation}
Q(x)\;=\;\langle x | Q\rangle
\end{equation}
$TQ$ equations (\ref{JQ}) take convenient form,
\begin{equation}\label{TQ}
\left\{
\begin{array}{l}
\ds t^{-1} A(u)  Q(x-\ii\bb) \;+\; t B(u) Q(x+\ii\bb)\;=\;T(u) Q(x)\;,\\
[5mm]
\ds [t^{-1}A(q^2u)]^* Q(x-\ii\bb^{-1}) \;+\; [tB(\frac{u}{q^2})]^* Q(x+\ii\bb^{-1}) \;=\; T(u)^* Q(x)\;.
\end{array}\right.
\end{equation}
Recall the convention \ref{star-convention}: for the real $x, \alpha_\nu,\beta_\nu$, the anti-homomorphism ``$*$'' coincides with the Hermitian conjugation "$\dagger$". 
However, one must consider the analytical continuation ``$*$'' rather then ``$\dagger$'' in the pair (\ref{TQ}) for general complex $x,\alpha_\nu,\beta_\nu$.

\section{Holomorphic and anti-holomorphic solutions.}

In this section we introduce the main building blocks for the solution to (\ref{TQ}).

Let holomorphic and anti-holomorphic functions\footnote{$\chi_{\pm,n}\sim q^{n(n+1)}$.}
\begin{equation}
\ch_+(u)\;=\;1+\sum_{n=1}^\infty \chi_{+,n} u^n\quad \textrm{and}\quad \ch_-(u)\;=\;1+\sum_{n=1}^\infty \chi_{-,n} u^{-n}
\end{equation}
be the solutions of
\begin{equation}\label{chi1}
\ch_+(\frac{u}{q^2}) \;+\; t^{2} A(q^2u)B(u) \ch_+(q^2u)\;=\;t T(u)\,\ch_+(u) 
\end{equation}
and 
\begin{equation}\label{chi2}
\ch_-(q^2u)  \;+\; t^2 A'(\frac{1}{u}) B'(\frac{q^2}{u})\, \ch_-(\frac{u}{q^2}) \;=\; \frac{tT(u)}{(-u)^N} \,\ch_-(u)\;.
\end{equation}
Their $q$-difference Wronskian
\begin{equation}\label{W}
W(u)\;=\;\ch_+(\frac{u}{q^2}) \ch_-(u) \;-\; t^2 (-qa)^N A(u) B'(\frac{q^2}{u}) \ch_+(u) \ch_-(\frac{u}{q^2})
\end{equation}
satisfies due to (\ref{chi1},\ref{chi2})
\begin{equation}\label{Wshift}
W(u)\;=\;(-u)^N W(q^2u)\;.
\end{equation}
Thus (see (\ref{theta})), one can introduce the ``Bethe Ansatz'' variables $x_\nu$ (equivalently $u_\nu$),
\begin{equation}\label{Wdecomp}
W(u)\;=\;\varrho\,\prod_{\nu=1}^N\, \vartheta_1\left(\frac{u}{u_\nu}\right)\;, \quad u_\nu\;=\;\EXP^{2\pi\bb x_\nu}\;, \quad \sum_{\nu=1}^N x_\nu\;=\;0\;.
\end{equation}
Presumably, the set  $x_\nu\in\mathbb{R}$ form a continuous distribution for a ground state of the transfer matrix $T(u)$ in the thermodynamic limit $N\to\infty$.

\section{Eigenfunction $Q(x)$ and spectral equations.}\label{Section5}

Solution $Q(x)$ of (\ref{TQ}) is given by
\begin{equation}
Q(x)\;=\;Q_1(x)\;-\;\xi\,Q_2(x)\;,
\end{equation}
where
\begin{equation}
Q_1(x)\;=\;\exp\left(-\,\sum_{\nu=1}^N \log\varphi(x-\alpha_\nu)\right) \, \frac{\ds \ch_+(u) \ch_-(u)^*}{\ds W(u)^*}\;,
\end{equation}
and
\begin{equation}
Q_2(x)\;=\;\exp\left(2\pi\ii\tau'xN - \sum_{\nu=1}^N \log\varphi(\beta_\nu-x)\right)\,
\frac{\ds \ch_-(u) \ch_+(u)^*}{\ds W(q^2u)^*}\;.
\end{equation}
Here we assume
\begin{equation}
t\;=\;\EXP^{2\pi\bb\tau}\;,\quad \textrm{and}\quad
\tau'\;=\;\frac{2\tau}{N}+\mu+\ii\eta\;.
\end{equation}
\begin{statement}
$Q_{1,2}(x)$ is the basis in the space of solutions of (\ref{TQ}) with proper asymptotic. Double-periodic coefficients are neglected as unwanted.
\end{statement}

\bigskip

\noindent
Now one can formulate the principle of quantisation of $T(u)$. The wave-function $Q(x)$ must have only ``kinematic'' poles coming from the dilogarithms $1/\varphi(x-\alpha_\nu)$ and $1/\varphi(\beta_\nu-x)$, i.e. the zeros of $W(u)^*$ must be canceled out:
\begin{equation}\label{BAE}
\lim_{x\to x_\nu}\frac{Q_1(x)}{Q_2(x)}\;=\;\xi\;,\quad \forall \;\;\nu=1,\dots, N\;.
\end{equation}
Definiton of the Wronskian (\ref{W}) guarantees that equations (\ref{BAE}) are then satisfied for all
\begin{equation}
x\;\to\;x_\nu + \ii\bb n - \ii\bb^{-1}m \;,\quad \forall \;\;n,m\in\mathbb{Z}\;.
\end{equation}

\section{Properties of $\ch_{\pm}$ for finite $N$.}

The principal system of the spectral equations is the system (\ref{BAE}). Initial variables are $\{x_\nu\}$ (equivalently $\{u_\nu\}$). However, the system (\ref{BAE}) involves the functions $\ch_{\pm}(u)$ being understood as the functions of the formal set of $\{u_\nu\}$:
\begin{equation}
\ch_{\pm}(u)\;=\;\ch_{\pm}(u; \{u_\nu\})\;.
\end{equation}
The main technical problem now is to reconstruct the functions $\ch_{\pm}(u)$ via $\{u_\nu\}$. 

For the finite $N$ one can use the $t^2$-expansion approach.
Namely, let
\begin{equation}\label{T0}
T_0(u)\;=\;\prod_{\nu=1}^N \left( 1\,-\,\frac{u}{u_\nu}\right)\;,\quad
T_0(u;q^2)_\infty \;=\;\prod_{\nu=1}^N\, \left(\frac{u}{u_\nu};q^2\right)_\infty\;,
\end{equation}
There is the $t^2$ expansion: 
\begin{equation}
\ch_+(u)\;=\;\sum_{m=0}^\infty t^{2m} F_m(u) T_0(q^{2(m+1)}u;q^2)_\infty\;,
\end{equation}
and
\begin{equation}
tT(u)\;=\;T_0(u)\;+\;\sum_{m=1}^\infty t^{2m} T_m(u)\;,
\end{equation}
where
\begin{equation}
F_0(u)\;=\;1\;,\quad F_m(u)\;=\;F_{m,1}u\;+\;\cdots\;+F_{m,mN}u^{mN}\;,
\end{equation}
so that $\ch_+(0)\;=\;1$, and
\begin{equation}
T_1(u)\;=\;1+\cdots+(-u)^N\;,\quad
T_m(u)\;=\; T_{m,1}u + \cdots + T_{m,N-1}u^{N-1}\;,
\end{equation}
in accordance with (\ref{T0N}). The $t^2$-expansion of equation (\ref{chi1}) gives
\begin{equation}\label{tpert}
\begin{array}{l}
\ds 
T_0(q^{2m}u) F_m(\frac{u}{q^2}) \;+\; A(q^2u)B(u) F_{m-1}(q^2u)
 \;=\;\\
[5mm]
\ds 
=\;\sum_{k=0}^m T_k(u) F_{m-k}(u) T_0(q^{2(m-k+1)}u)\cdots T_0(q^{2m}u)\;.
\end{array}
\end{equation}
\begin{statement}\label{existence}
The system (\ref{tpert}) provides a linear bootstrap allowing one to construct uniquely step by step all $F_m(u)$ and $T_m(u)$.
\end{statement}
\noindent
The bootstrap (\ref{tpert}) involves the finite polynomials and therefore it can be solved either straightforwardly or by a partial fraction decomposition. However, we failed to find any manageable expression for $F_m(u)$ and $T_m(u)$ except for the simplest case $N=1$ (see below). Thus, the statement \ref{existence} can be seen yet as a ``theorem of existence''.

\bigskip

\noindent
Consider the zero decomposition of $\ch_{\pm}(u)$:
\begin{equation}\label{chiprod}
\ch_+(u)\;=\;\prod_{m=1}^\infty \prod_{\nu=1}^N \left(1-q^{2m}\frac{u}{u_{\nu,m}}\right)\;,\quad
\ch_-(u)\;=\;\prod_{m=1}^\infty \prod_{\nu=1}^N \left(1-q^{2m}\frac{u_{\nu,m}'}{u}\right)\;.
\end{equation}
It is easy to show, the coefficients $u_{\nu,m}$ entering the zero decomposition (\ref{chiprod}) have the following asymptotic:
\begin{equation}\label{as1}
\frac{u_\gamma}{u_{\gamma,n}}\;=\;1\;+\;t^{2n}\Delta_{\gamma,n}'\;+\;\mathcal{O}(t^{2n+2})\;,
\end{equation}
where
\begin{equation}\label{Delta1}
\Delta_{\gamma,n}'\;=\;\lim_{u\to u_\gamma} \left(1-\frac{u_\gamma}{u}\right) \Delta'(1/u;q^2)_n\;,\quad 
\Delta'(1/u;q^2)_n\;=\;
\frac
{\ds A'(1/u;q^2)_n B'(q^2/u;q^2)_n}
{\ds T_0'(1/u;q^2)_n T_0'(q^2/u;q^2)_n}\;.
\end{equation}
Here in addition to (\ref{T0}) we use
\begin{equation}
T_0'(1/u) \;=\; \frac{T_0(u)}{(-u)^N}\;=\;\prod_{\nu=1}^N \left( 1 - \frac{u_\nu}{u}\right)\;,\quad 
T'_0(1/u;q^2)_n\;=\;\prod_{\nu=1}^N \left(\frac{u_\nu}{u};q^2\right)_n\;.
\end{equation}
In the similar way
\begin{equation}\label{as2}
\frac{u_{\gamma,n}'}{u_{\gamma}}\;=\;1\;+\;t^{2n}\Delta_{\gamma,n}\;+\;\mathcal{O}(t^{2n+2})\;,
\end{equation}
where
\begin{equation}\label{Delta2}
\Delta_{\gamma,n}\;=\;\lim_{u\to u_\gamma} \left(1-\frac{u}{u_\gamma}\right) \Delta(u;q^2)_n\;,\quad
\Delta(u;q^2)_n\;=\;
\frac
{\ds A(q^2u;q^2)_n B(u;q^2)_n}
{\ds T_0(u;q^2)_n T_0(q^2u;q^2)_n}\;.
\end{equation}

\subsection{Remarks}

Also we'd like to point out remarkable symmetric case: choosing
\begin{equation}
\beta_\nu\;=\;-\alpha_\nu
\quad \textrm{and} \quad 
x_\nu\;=\;-x_{-\nu}\;,
\end{equation}
one gets 
\begin{equation}
T_0'(1/u)\;=\;T_0^{}(1/u)\;,\quad \frac{T(u)}{(-u)^N}\;=\;T(1/u)\;,\quad \ch_-(u)\;=\;\ch_+(1/u)\;.
\end{equation}
In this case the parameter $\xi$ in (\ref{BAE}) becomes the parity of the state, $\xi=\pm 1$:
\begin{equation}
\EXP^{-2\pi\ii\tau' x_\gamma N} \;\left(\prod_{\nu} \frac{\varphi(-\alpha_\nu-x_\gamma)}{\varphi(x_\gamma-\alpha_\nu)}\right)\;
\frac{\ch_+(u_\gamma) \ch_+(1/u_\gamma)^*}{\ch_+(1/u_\gamma) \ch_+(u_\gamma)^*}\;=\;\pm 1\;,\quad
\gamma=1,\cdots,\frac{N}{2}\;.
\end{equation}

\subsection{The toy case $N=1$}

When $N=1$, then the $t^2$-series expansion is manageable:
\begin{equation}
\ch_+(u)\;=\;\sum_{m=0}^\infty
q^{m(m+1)} \;
\frac{ \ds (-t^2u)^m }{ \ds (q^2;q^2)_m} \;
\frac{ \ds (-qa;q^2)_m (-qb;q^2)_m }{\ds (q^2t^2;q^2)_m }\; 
(q^{2(m+1)}u;q^2)_\infty\;,
\end{equation}
and 
\begin{equation}
\ch_-(u)\;=\;\ch_+(1/u)\;.
\end{equation}
Transfer-matrices are
\begin{equation}
T_0(u)\;=\;T_1(u)\;=\;1-u\;,\quad \textrm{and}\;\;\;T_m(u)\;=\;0\quad \textrm{for}\;\;m\geq 2\;.
\end{equation}
The $q$-Wronskian (\ref{W}) is then
\begin{equation}
W(u)\;=\; \frac{\ds (-q t^2 a;q^2)_\infty (-qt^2b;q^2)_\infty}{\ds (q^2t^2;q^2)_\infty^2}\; \vartheta_1(u)\;.
\end{equation}

\section{Thermodynamical limit.}

So far we do not have essential numerical results in finding $u_{\gamma,m}$ via $u_\gamma$, however we have an irresistible temptation to make some traditional assumptions for the thermodynamic limit $N\to\infty$. 

First of all, for the limit $N\to\infty$ define the ``external'' densities $P_A$ and $P_B$:
\begin{equation}
\frac{1}{N}\sum_\nu f(\alpha_\nu)\;=\;\int dx_0 P_A(x_0) f(x_0)\;,\quad 
\frac{1}{N}\sum_\nu f(\beta_\nu)\;=\;\int dx_0 P_B(x_0) f(x_0)\;.
\end{equation}
Expectations for $P_A$ and $P_B$ are
\begin{equation}
\mu_A\;=\;\int dx_0 P_A(x_0) x_0\;=\;\mu\;,\quad 
\mu_B\;=\;\int dx_0 P_B(x_0) x_0\;=\;-\mu\;,
\end{equation}
see (\ref{abnormalization}). Also, let
\begin{equation}
S_A\;=\;\int dx_0 P_A(x_0) x_0^2\;,\quad 
S_B\;=\;\int dx_0 P_B(x_0) x_0^2\;.
\end{equation}
The symmetric case implies $P_B(x_0)=P_A(-x_0)$, $S_A=S_B$. 
The homogeneous symmetric case corresponds fo $P_A(x_0)\;=\;\delta(x_0-\mu)$, $P_B(x_0)\;=\;\delta(x_0+\mu)$.

\bigskip

\noindent
In what follows, let for shortness
\begin{equation}
P_{AB}(x_0) \;=\; P_A(x_0) + P_B(x_0)\;.
\end{equation}

\bigskip

\noindent
Next we assume a continuous distribution of the Bethe-type variables $x_\nu$ on the real axis. Let their density be an analytical function $P(x_0)$, so that e.g.
\begin{equation}
\frac{1}{N}\log T_0(u)\;=\;\int_{\mathbb{R}\pm\ii 0} dx_0 P(x_0) \log (1-\EXP^{2\pi\bb (x-x_0)})\;,\quad x\in\mathbb{R}\;,
\end{equation}
where the notation ``$\mathbb{R}\pm \ii 0$'' marks a proper choice of the branch of the logarithm.

\bigskip
\noindent
Condition $x\in \mathbb{R}$ is implied everywhere below.

\bigskip

\noindent
Next assumption is that the roots $x_{\gamma,m}$ are continuously distributed along some complex contours $\mathcal{C}_m$, however these contours can be analytically straightened to the real axis with the same distribution density $P(x_0)$ as before. Later we will see that there is even more strong behaviour:
\begin{equation}\label{mlimit}
x_{\gamma,m}\;\to\; x_\gamma\quad \forall\;\;m\quad \textrm{when}\;\; N\to 0\;.
\end{equation}
This is based on
\begin{equation}\label{Delta0}
\frac{1}{N}\log \left|\Delta_{\gamma,n}\right| \;<\; 0\;,
\end{equation}
where $\Delta_{\gamma,n}$ enters the asymptotic of $x_{\gamma,m}$, see (\ref{Delta1},\ref{Delta2}).

\bigskip

\noindent
With the assumptions made, the left hand side of Bete Ansatz equations (\ref{BAE}) become
\begin{equation}\label{QQ1}
\begin{array}{l}
\ds \frac{1}{N}\log\frac{Q_1(x)}{Q_2(x)}\;=\; -2\pi\ii (\mu+\ii\eta) x + 
\int dx_0 P(x_0) \log \frac{\varphi(x-x_0+\ii\eta)}{\varphi(x_0-x+\ii\eta)}\\
[5mm]
\ds -\int dx_0 P_A(x_0) \log\varphi(x-x_0) + \int dx_0 P_B(x_0)\log\varphi(x_0-x) \;.
\end{array}
\end{equation}
Elementary transformations provide
\begin{equation}\label{QQ2}
\begin{array}{l}
\ds \frac{1}{N}\log\frac{Q_1(x)}{Q_2(x)}\;=\;\ii\pi (\eta^2+S_B-S_P)\\
[5mm]
\ds 
+ \int dx_0 (P(x_0+\ii\eta)+P(x_0-\ii\eta) -P_{AB}(x_0))\log\varphi(x-x_0)\;,
\end{array}
\end{equation}
where the normalisation and expectations
\begin{equation}
\int dx_0^{} P(x_0^{})\;=\;1\;,\quad \int dx_0^{} P(x_0^{}) x_0^{} \;=\;0\;,\quad \int dx_0^{} P(x_0^{}) x_0^2 \;=\; S_P\;.
\end{equation}
are taken into account. Thus, the Bethe Ansatz equations (\ref{BAE}) in the integral form (\ref{QQ2}) give 
\begin{equation}\label{primeq}
P(x_0+\ii\eta)+P(x_0-\ii\eta)\;=\;P_{A}(x_0)+P_B(x_0)
\end{equation}
for the ground state density $P(x_0)$. On the solution of (\ref{primeq}) one has $\ds S_P = \eta^2 + \frac{1}{2}(S_A+S_B)$, so that the Bethe Ansatz equations (\ref{BAE}) become the identity 
\begin{equation}
\frac{1}{N}\log \frac{Q_1(x)}{Q_2(x)}\;=\;\frac{\ii\pi}{2} (S_B-S_A)
\end{equation}

\bigskip

\noindent
Solution to (\ref{primeq}) gives the ground state density
\begin{equation}\label{PBethe}
P(x)\;=\;\int dx_0 K(x-x_0) P_{AB}(x_0)\;,
\end{equation}
where the kernel
\begin{equation}
K(x)\;=\;
\int \frac{\ds \EXP^{-2\pi\ii xy}}{\ds 2\cosh(2\pi\eta y)} dy\;=\;
\frac{1}{\ds 4\eta \cosh\left( \frac{\pi x}{2\eta} \right)}
\;.
\end{equation}
In particular, in the homogeneous symmetric case $P_{AB}(x)=\delta(x-\mu)+\delta(x+\mu)$,
\begin{equation}\label{distribution}
P(x)\;=\;
\frac{1}{\ds 4\eta \cosh\left( \frac{\pi (x-\mu)}{2\eta} \right)}
+
\frac{1}{\ds 4\eta \cosh\left( \frac{\pi (x+\mu)}{2\eta} \right)}\;.
\end{equation}

\bigskip
\noindent
Expressions for $\ch_{\pm}(u)$ then become
\begin{equation}\label{chch}
\frac{1}{N}\log \frac{\ch_+(u)\ch_-(u)^*}{W(u)^*}\;=\;\Phi_1(x) \;,\quad 
\frac{1}{N}\log \frac{\ch_-(u)\ch_+(u)^*}{W(q^2u)^*}\;=\;\Phi_2(x)\;,
\end{equation}
where
\begin{equation}\label{Phi1}
\Phi_1(x) \;=\; \int dx_0 P(x_0) \log\varphi(x-x_0+\ii\eta)\;=\;\int dx_0 P_{AB}(x_0) \log\varphi_2(x-x_0+\ii\eta)\;,
\end{equation}
\begin{equation}\label{Phi2}
\Phi_2(x)\;=\; \int dx_0 P(x_0) \log\varphi(x_0-x+\ii\eta)\;=\;\int dx_0 P_{AB}(x_0) \log\varphi_2(x_0-x+\ii\eta)\;.
\end{equation}
Expressions (\ref{Phi1}) and (\ref{Phi2}) with the homogeneous density (\ref{distribution}) being analytically continued to the stat mechanical regime of imaginary $x$ and $\mu$ explicitly reproduce the free energy for the Faddeev-Volkov model \cite{Bazhanov_2008}. Moreover, for imaginary $\mu$ the distribution (\ref{distribution}) becomes unimodal.

\bigskip
\noindent
The transfer matrix is given by
\begin{equation}\label{logt}
\frac{1}{N} \log T_0(u) \;=\; \int_{\mathbb{R}\pm\ii 0} dx_0 P(x_0) \log(1-\EXP^{2\pi\bb (x-x_0)})\;=\; \int dx_0 P_{AB}(x_0) I_\pm(x-x_0)\;,
\end{equation}
where
\begin{equation}\label{I}
I_{+}(x)\;=\;\log \frac{(-qu;q^4)_\infty}{(-q^3u;q^4)_\infty} \frac{(-\ii w; -\ii p)_\infty}{(\ii w; -\ii p)_\infty}
\end{equation}
with
\begin{equation}
w\;=\;\exp \left(\frac{\pi x}{2\eta}\right)\;,\quad p\;=\;\exp \left( -\frac{\pi\sigma}{2\eta} \right)\;,\quad -\ii p\;=\;\exp \left(-\frac{\ii\pi\bb^{-1}}{2\eta} \right)\;,
\end{equation}
and 
\begin{equation}
I_{-}(x) \;=\; I_+(x) - 2\pi\ii y(x)\;,\quad y(x)\;=\;\int_{-\infty}^x dx_0 K(x_0)\;=\;\frac{1}{2\pi\ii}\log\frac{1+\ii w}{1-\ii w}\;.
\end{equation}
These computations allow one to verify condition (\ref{Delta0}):
\begin{equation}
\frac{1}{N}\log \left|\frac{A(q^2u) B(u)}{T_0(u) T_0(q^2 u)}\right|\;=\;\int dx_0 P_{AB}(x_0) \log\tanh\left|\frac{\pi(x-x_0-\sigma)}{4\eta}\right|\;<\;0
\end{equation}
Note that due to the branch cuts of $\log$ in (\ref{logt}) the expression for $I(x)$ must be modified in each strip $k\eta < \textrm{Im}\; x < (k+1)\eta$, $x\in\mathbb{C}$. In particular, $\log T(u)$ has no singularities.
Note also,
\begin{equation}
\frac{1}{N}\log\left|\frac{A(q^2u)B(u)}{ T_0(u) T_0(u)_{x\to x+2\ii\eta}}\right| \;=\; 0\;,
\end{equation}
where $\log T_0(u)_{x\to x+2\ii\eta}$ corresponds to the analytical continuation $I_{\pm}(x)\to I_{\pm}(x+2\ii\eta)$ in  (\ref{logt}).

\bigskip
\noindent
The results here correspond to the ground state structure of the model. Next step is the study of the low energy excitations. 

\bigskip
\noindent
\textbf{Acknowledgements.} I would like to thank R. Kashaev, V, Bazhanov and V. Mangazeev for valuable discussions. Also I acknowledge the support of the Australian Research Council grant 
DP190103144.


\begin{thebibliography}{10}

\bibitem{Faddeev_1995}
L.~D. Faddeev.
\newblock Discrete Heisenberg-Weyl group and modular group.
\newblock {\em Letters in Mathematical Physics}, Vol. 34(3) pages 249-254, 1995.

\bibitem{Faddeev_1999}
L.~Faddeev.
\newblock Modular double of a quantum group.
\newblock {\em In Conference Mosh`e Flato 1999, Vol. I (Dijon), volume 21
  of Math. Phys. Stud, pages 149-156, Dordrecht, 2000. Kluwer Acad. Publ.},
  1999.

\bibitem{Sergeev_2005}
S.~M. Sergeev.
\newblock Quantization scheme for modular $q$-difference equations.
\newblock {\em Theoretical and Mathematical Physics}, Vol 142(3), pages 422-430, 2005.

\bibitem{Kashaev_2018}
Rinat~M. Kashaev and Sergey~M. Sergeev.
\newblock Spectral equations for the modular oscillator.
\newblock {\em Reviews in Mathematical Physics}, Vol 30(07), page1840009, 2018.

\bibitem{Babelon_2018}
Olivier Babelon, Karol~K. Kozlowski, and Vincent Pasquier.
\newblock Baxter operator and Baxter equation for $q$-toda and toda2 chains.
\newblock {\em Reviews in Mathematical Physics}, Vol 30(06), page 1840003, 2018.

\bibitem{Kashaev_2020}
Rinat Kashaev and Sergey Sergeev.
\newblock On the spectrum of the local ${\mathbb {P}}^2$ mirror curve.
\newblock {\em Annales Henri Poincare}, Vol 21(11), pages 3479-3497, 2020.

\bibitem{Bazhanov_2007}
Vladimir~V. Bazhanov, Vladimir~V. Mangazeev, and Sergey~M. Sergeev.
\newblock Faddeev-Volkov solution of the Yang-Baxter equation and discrete conformal symmetry.
\newblock {\em Nuclear Physics B}, Vol 784(3), pages 234-58, 2007.

\bibitem{Lukyanov_1997}
Sergei Lukyanov and Alexander Zamolodchikov.
\newblock Exact expectation values of local fields in the quantum sine-Gordon model.
\newblock {\em Nuclear Physics B}, Vol 493(3), pages 571-587, 1997.

\bibitem{Faddeev_2001}
L.~D. Faddeev, R.~M. Kashaev, and A.~Yu. Volkov.
\newblock Strongly coupled quantum discrete Liouville theory I: Algebraic
  approach and duality.
\newblock {\em Communications in Mathematical Physics},  Vol 219(1), pages 199-219, 
  2001.

\bibitem{Bazhanov_2008}
Vladimir~V. Bazhanov, Vladimir~V. Mangazeev, and Sergey~M. Sergeev.
\newblock Exact solution of the Faddeev-Volkov model.
\newblock {\em Physics Letters A}, Vol 372(10), pages 1547-1550, 2008.

\end{thebibliography}

\end{document}